
\magnification 1200
\font\eightrm=cmr8 \font\eighti=cmmi8 \font\eightsy=cmsy8
\font\eightbf=cmbx8 \font\eighttt=cmtt8
\font\eightit=cmti8 \font\eightsl=cmsl8
\font\sixrm=cmr6 \font\sixi=cmmi6 \font\sixsy=cmsy6 \font\sixbf=cmbx6
\catcode`@11
\newskip\ttglue
\def\sqr#1#2{{\vcenter{\hrule height.#2pt   \hbox{\vrule width.#2pt
height#1pt \kern#1pt    \vrule width.#2pt}\hrule height.#2pt}}}

\def\Real{\hbox{\hskip 1.5pt\hbox to 0pt{\hskip -2pt I\hss}R}}
\def\Complex{\hbox{\hbox to 0pt{\hskip 2.7pt \vrule height 6.5pt
                                  depth -0.2pt width 0.8pt \hss}C}}
\centerline {\bf General Neveu-Schwarz Correlators in Super Liouville Theory}
\vskip 1cm
\centerline {E. Abdalla$^1$, M.C.B. Abdalla$^2$,}

\centerline {D. Dalmazi$^2$, Koji Harada$^{3}$}
\vskip .2cm
\centerline {$^1$Instituto de F\'\i sica, Univ. S\~ao Paulo, CP 20516,
S\~ao Paulo, Brazil}
\vskip .2cm

\centerline {$^2$Instituto de F\'\i sica Te\'orica, UNESP, Rua Pamplona 145,}

\centerline {CEP 01405, S\~ao Paulo, Brazil}
\vskip .2cm
\centerline {$^3$Department of Physics, Kyushu University, Fukuoka 812, Japan}
\vskip 2cm
\centerline {\bf Abstract}

In this paper we compute the N-point correlation functions of the tachyon
operator from
the Neveu Schwarz sector of super Liouville theory coupled to matter fields
 (with $\hat c\le 1$) in the super Coulomb gas formulation,
 on world sheets with spherical topology. We
first integrate over the zero mode
assuming that the $s$  parameter takes an integer value, subsequently
 we continue the parameter to an arbitrary real number. We included an
arbitrary
number of screening charges (s.c.) and  as a result, after  renormalizing the
s.c., the external legs and the cosmological constant, the form of the final
amplitudes  do not modify. Remarkably, the result is
completely parallel to the bosonic case. We also completed a discussion on the
calculation of bosonic correlators including arbitrary screening  charges.
\vfill \eject

\centerline {\bf  1- Introduction }
\vskip .50cm

\def\eightpoint{\def\rm{\fam0\eightrm}
\textfont0=\eightrm \scriptfont0=\sixrm \scriptscriptfont0=\fiverm
\textfont1=\eighti \scriptfont1=\sixi \scriptscriptfont1=\fivei
\textfont2=\eightsy \scriptfont2=\sixsy \scriptscriptfont2=\fivesy
\textfont3=\tenex \scriptfont3=\tenex \scriptscriptfont3=\tenex
\textfont\itfam=\eightit \def\it{\fam\itfam\eightit}
\textfont\slfam=\eightsl \def\sl{\fam\slfam\eightsl}
\textfont\ttfam=\eighttt \def\tt{\fam\ttfam\eighttt}
\textfont\bffam=\eightbf
\scriptfont\bffam=\sixbf
\scriptscriptfont\bffam=\fivebf \def\bf{\fam\bffam\eightbf}
\tt \ttglue=.5em plus.25em minus.15em
\normalbaselineskip=6pt
\setbox\strutbox=\hbox{\vrule height7pt width0pt depth2pt}
\let\sc=\sixrm \let\big=\eightbig \normalbaselines\rm}
\newinsert\footins
\def\newfoot#1{\let\@sf\empty
  \ifhmode\edef\@sf{\spacefactor\the\spacefactor}\fi
  #1\@sf\vfootnote{#1}}
\def\vfootnote#1{\insert\footins\bgroup\eightpoint
  \interlinepenalty\interfootnotelinepenalty
  \splittopskip\ht\strutbox 
  \splitmaxdepth\dp\strutbox \floatingpenalty\@MM
\leftskip\z@skip \rightskip\z@skip
\textindent{#1}\footstrut\futurelet\next\fo@t}
\def\fo@t{\ifcat\bgroup\noexpand\next \let\next\f@@t
  \else\let\next\f@t\fi \next}
\def\f@@t{\bgroup\aftergroup\@foot\let\next}
\def\f@t#1{#1\@foot}  \def\@foot{\strut\egroup}
\def\footstrut{\vbox to\splittopskip{}}
\skip\footins=\bigskipamount \count\footins=1000
\dimen\footins=8in 
\def\ref#1{$^{#1}$}
\newbox\bigstrutbox
\setbox\bigstrutbox=\hbox{\vrule height10pt depth5pt width0pt}
\def\bigstrut{\relax\ifmmode\copy\bigstrutbox\else\unhcopy\bigstrutbox\fi}

Matrix models proved to be an efficient means to obtain
information about
non-critical string theory, specially in computations of correlation
functions\ref{1}. However, supersymmetric extensions\ref{2} are,
up to now, not constructed, due to some technical difficulties.

On the other hand, in the continuum approach, we have
to deal with Liouville theory\ref{3}. Unfortunately, we still do not
understand it very well in spite of much effort. In particular, it has been
difficult to calculate
correlation functions in a reliable way because  perturbation theory does not
apply. Recently, however, several authors\ref{4-10} succeeded in taming the
difficulties of the Liouville theory and computed exactly correlation functions
in the continuum approach to conformal fields coupled to two dimensional
gravity. The technique is based on the integration over the zero mode of
the Liouville field. The resulting amplitude is a function  of a
parameter $s$ which
depends on the central charge and on the external momenta. The
amplitudes can be computed when the above parameter is a non-negative
integer.
Later on, one analytically continues that parameter to  real (or complex)
values. The results thus obtained agree with the matrix model approach.

An advantage of the Liouville approach is that it is very easy to
extend it to
supersymmetric theories\ref{12}. Since matrix models seem to be less
powerful (up to now) in this point, it is natural to investigate supersymmetric
Liouville
theory in order to know something about 2-D supergravity (or off-critical
NSR string).

Our aim here is to investigate the supersymmetric Liouville theory.
We shall
compute supersymmetric correlation functions on world sheets with
spherical
topology in the Neveu-Schwarz sector, where the super-Liouville is
coupled
to superconformal matter with central charge $\hat c\le 1$, represented
as a super
Coulomb gas\ref{12,13}. The results are remarkable, and very parallel to the
bosonic case;
it is equivalent to a redefinition of the cosmological constant,
and of the
primary superfields, the resulting  amplitudes have the same form as those of
the bosonic theory obtained by Di Francesco and Kutasov\ref{5}. The super
Liouville theory\ref{14} has also been studied along similar lines. Our present
results generalize those presented in a recent paper\ref{15}, as well
as others recently obtained in the literature\ref{16}.

This paper is divided as follows: in section 2 we review the computations of
bosonic correlators, and
include the case with an arbitrary number of screening charges.
 In section 3 we calculate the N-point tachyon (Neveu-Schwarz) correlation
functions also with screening charges. Results are
remarkably similar to the bosonic case\ref{5}. In section 4 we draw some
conclusions. In Appendix A, we calculate the supersymmetric generalization of
equation (B.10) of Dotsenko and Fatteev.

\vskip 1cm
\penalty-200
\centerline {\bf 2- Bosonic Correlators}
\nobreak
\vskip .5cm
\nobreak
\noindent {\bf 2.1- The 3-point tachyon amplitude}
\vskip .5cm
\nobreak
In a recent paper Di Francesco and Kutasov\ref{6} calculated
the $N$-point tachyon correlation functions in Liouville theory on
world sheets with spherical topology, coupled to $c\le 1$ conformal
matter in a Coulomb gas
representation. They worked in the DDK's framework\ref{17} where
the total action
is given by:
$$
S={1\over 2\pi} \int d^2w \sqrt {\hat g}
\left[ \hat {g^{ab}}\partial_a\phi\partial _b\phi
-{Q\over 4}\hat R \phi
+2\mu e^{\alpha \phi} + {\hat g}^{ab} \partial _aX\partial _bX
+{i\alpha_0\over2}\hat R X\right]\quad , \eqno(2.1)
$$
where $\phi$ is the Liouville mode and $X$ represents the matter with
$c=1-12\alpha_0^2$. Following DDK, the constant $Q$ is determined
by imposing a vanishing total central charge,
$$
Q=\sqrt {{25-c\over 3}}=2\sqrt {2+\alpha_0^2}\quad ,\eqno(2.2)
$$
and $\alpha$ is determined by requiring $e^{\alpha\phi}$ to be
a (1,1) conformal operator,i.e., $-{1\over 2}\alpha(\alpha+Q)=1$.
We define the solutions to this equation  by:
$$
\alpha_\pm=-{Q\over 2}\pm \vert \alpha_0\vert \quad ,\quad
\alpha_+\alpha_-=2\quad ;\eqno(2.3)
$$
the semiclassical limit $(c \to \infty)$ fixes $\alpha =\alpha_+$.

In the following we calculate gravitationally-dressed tachyon
amplitudes:
$$
\langle T_{k_1}\cdots T_{k_N}\rangle =
\left\langle
\prod_{j=1}^N\int d^2z_je^{ik_jX(z_j)+\beta(k_j)\phi(z_j)}
\right\rangle
\eqno(2.4)
$$
the dressing parameter $\beta$ is fixed imposing  $e^{ik_jX+
\beta_j\phi}$ to be a (1,1) conformal operator:
$$
\beta_j=\beta (k_j)=-{Q\over 2} +\vert k_j-\alpha_0\vert
\quad .\eqno(2.5)
$$

An important ingredient in the calculation of $\langle T_{k_1}\cdots
T_{k_n}\rangle $ is the integration over the matter $(X_0)$ and
the Liouville
$(\phi_0)$ zero modes. We make the following split\ref{4,5}
$$
\eqalign{
\phi&=\phi_0+\tilde \phi \cr
X&=X_0+\tilde X \cr
}\eqno(2.6)
$$
where the fields $\tilde \phi$ and $\tilde X$ obey the condition
$$
\int d^2w\tilde \phi =\int d^2w \tilde X=0\quad .\eqno(2.7)
$$
The integration over the zero modes   $X_0$ and $\phi_0$  gives the
following
results;
$$
\int _{-\infty}^{\infty}{\cal D}X_0e^{iX_0\left(\sum_{i=1}^Nk_i
-2\alpha_0\right) }
=2\pi \delta( \sum_{i=1}^Nk_i-2\alpha_0 )\quad,
\eqno(2.8a)
$$
$$
\int _{-\infty}^{\infty}{\cal D}\phi_0
e^{
i\phi_0\left(\sum_{j=1}^N\beta_j+Q\right)
-e^{\alpha_+\phi_0\left( {\mu \over \pi}\int d^2w
e^{\alpha_+\tilde\phi}\right)}
} ={\Gamma (-s)\over -\alpha_+}
\left( {\mu\over \pi }\int d^2w e^{\alpha_+\tilde\phi}\right) ^s
\quad ,\eqno(2.8b)
$$
where
$$
s=-{1\over \alpha_+}\left( \sum _{j=1}^N\beta_j+Q\right)
\quad .\eqno(2.9)
$$
And we have used that on the sphere ${1\over 8\pi} \int d^2 w \sqrt
{\hat g}\hat R=1$.
We thus obtain
$$
\langle T_{k_1}\cdots T_{k_N}\rangle =2\pi \delta\left(
\sum_{j=1}^N k_j-2
\alpha_0 \right) {\cal A}_N(k_1\cdots k_N)
\eqno(2.10a)
$$
$$
{\cal A}_N(k_1\cdots k_N)={\Gamma (-s)\over -\alpha_+}
\left( {\mu \over \pi
}\right) ^s\left\langle \prod_{j=1}^N\int d^2z_j
e^{ik_j+\beta_j\phi(z_j)}
\left( \int d^2w e^{\alpha_+\phi}\right) ^s\right\rangle_0
\eqno(2.10b)
$$
where $\langle \cdots \rangle _0$ means that now the correlation
functions are
calculated as in the free theory $(\mu =0)$. The strategy to obtain
${\cal A}_N$
is to assume first that $s$ is a non-negative integer and to continue the
result to any real  $s$ at the end. Thus, using free propagators:\newfoot{$^{1)
}$}{Hereafter we drop the tilde in the fields defined by (2.9), since no
confusion can occur.}
$$
\langle X(w)X(z)\rangle_0=\langle\phi(w)\phi(z)\rangle_0
= \ln \vert w-z\vert ^{-2}
\eqno(2.11)
$$
and fixing the residual
$SL(2\Complex )$ invariance of the conformal gauge on the sphere
by choosing
$(z_1=0\, ,\, z_2=1\, ,\, z_3=\infty)$, we have in the case of the 3-point
function:
$$
{\cal A}_3(k_1,k_2,k_3)={\Gamma (-s)\over -\alpha_+}
\left( {\mu \over \pi}\right) ^s\int \prod_{j=1}^sd^2w_j
\vert w_j\vert ^{2\alpha} \vert 1-w_j\vert ^{2\beta}
\prod_{i<j}^s\vert w_i-w_j\vert ^{4\rho}
\quad .\eqno(2.12)
$$

Above, we defined  $\alpha=-\alpha_+\beta_1\, ,\, \beta=-\alpha_+\beta_2\, ,\,
\rho=-\alpha_+^2/2$.  The above integral has
been calculated by Dotsenko and Fatteev\ref{12} ( see formula (B.9), of the
 second  paper). Using their result Di Francesco and Kutasov obtained:
$$
\eqalign{
{\cal A}_3(k_1,k_2,k_3)&={\Gamma (-s)\over -\alpha_+}
\left( {\mu \over \pi}\right) ^s\Gamma(s+1)[\Delta(1-\rho)]^s
\prod_{i=1}^s\Delta(i\rho)\cr
&\times
\prod_{i=0}^{s-1}\Delta(1+\alpha+i\rho)\Delta(1+\beta+i\rho)
\Delta(-1-\alpha-\beta-(s-1+i)\rho)\cr }
\eqno(2.13)
$$
where $\Delta(x)=\Gamma(x)/\Gamma(1-x)$. Choosing the
kinematics\newfoot{$^{2)}$}{Notice that our notation differs from Ref.[6] by
the exchange of $k_2$ and  $k_3$.}
$k_1,k_3\ge
\alpha_0 \, ,\, k_2<\alpha_0$ we can eliminate $\beta$ using
(2.6), (2.9) and the
momentum conservation: $\sum_{i=1}^3k_i=2\alpha_0$:,
$$
\beta=\cases{\rho(1-s)\quad ,\quad \alpha_0>0\cr
-1-\rho\quad ,\quad \alpha_0<0\cr}
\eqno(2.14)
$$
Back in (2.13) it is easy to see that for $\alpha_0>0$ there
appears a factor
$\Gamma(0)$ in the denominator of ${\cal A}_3$ and the amplitude
vanishes
identically. For $\alpha_0<0$, using
$$
\eqalignno{
{1\over 2}(\beta_1^2-k_1^2)&=\rho-\alpha\quad ,&(2.15a)\cr
{1\over 2}(\beta_2^2-k_2^2)&=1+\alpha (s-1)\rho \quad ,&(2.15b)\cr
{1\over 2}(\beta_3^2-k_3^2)&=-s \quad ,&(2.15c) \cr}
$$
we can write the 3-point amplitude in a rather compact form
$$
{\cal A}_3=
[\mu \Delta(-\rho)]^s\prod_{j=1}^3\Delta \left( {1\over2}
(\beta_j^2-k_j^2)\right)
\quad . \eqno(2.16)
$$
Thus, after the redefinitions of the cosmological constant and
of the external
fields as
$$
\mu \to {\mu \over \Delta(-\rho)}\quad ,
\quad T_{k_j}\to {T_{k_j}\over \Delta
\left( {1\over 2}(\beta_j^2-k_j^2)\right)}\quad ,
\eqno(2.17)
$$
Di Francesco and Kutasov\ref{6} obtained  for the three-point function
$$
{\cal A}_3=\mu ^s \quad ,\eqno(2.18)
$$
which is also obtained in the matrix model approach. In the next
sub-sections we
shall see that this expression holds for general N-point tachyon amplitudes
with an arbitrary number of s.c..

\vskip 1cm
\noindent {\bf 2.2- The 3-point tachyon amplitude with an arbitrary
number of screening charges }
\vskip 1cm

Now we show explicitly how one generalizes the previous
calculation to the case which includes an arbitrary number of
screening charges in the matter sector. we introduce $n$ operators
$e^{id_+X}$ and $m$ operators $e^{id_-X}$,
with $d_\pm$ solutions of: ${1\over 2}d(d-2\alpha_0)=1\, ,\,
(d_+d_-=-\alpha_+\alpha_- =-2)$. Integrating over the zero-modes
again we get:
$$
\eqalign{
&\left\langle T_{k_1}T_{k_2}T_{k_3}
\left( {1\over n!}\prod _{i=1}^n\int d^2t_ie^{id_+X(t_i)}\right)
\left( {1\over m!}\prod _{i=1}^m\int d^2r_ie^{id_-X(r_i)}\right)
\right\rangle \cr
= &2\pi \delta(\sum _{i=1}k_i+nd_++md_--2\alpha_0)
{\cal A}^{nm}_3(k_1,k_2,k_3)
\cr}\eqno(2.19)
$$
where the amplitude ${\cal A}^{nm}_3(k_1,k_2,k_3)$ is given by
the expression
$$\eqalign{
{\cal A}_3^{nm}(k_1,k_2,k_3)&
={\Gamma(-s)\over -\alpha_+}\left( {\mu \over \pi}\right) ^s
\prod _{i=1}^n\int d^2t_i\vert t_i\vert ^{2\tilde \alpha}
\vert 1-t_i\vert^{2\tilde \beta}
\prod _{i<j}^n\vert t_i-t_j\vert ^{4\tilde \rho}\cr
&\times \prod _{i=1}^m\int d^2r_i\vert r_i\vert^{2\tilde \alpha'}
\vert 1-r_i\vert^{2\tilde \beta'}
\prod_ {i<j}^m\vert r_i-r_j\vert ^{4\tilde\rho'}\cr
&\times \prod_{i=1}^{n}\prod_{j=1}^m\vert t_i-r_j\vert^{-4}
\prod_{i=1}^s\int d^2z_i\vert z_i\vert ^{2\alpha}
\vert 1-z_i\vert^{2\beta}\prod_{i<j}^s\vert z_i-z_j\vert ^{4\rho}\cr}
\eqno(2.20)
$$
The parameters  $\alpha,\beta$ and $\rho$ are defined as before,
and the remaining parameters  are
$$\eqalign{
\tilde  \alpha &=d_+k_1\quad ,\quad \tilde  \beta = d_+k_2\quad ,\quad
\tilde  \rho ={1\over 2}d_+^2 \cr
\tilde  \alpha '&=d_-k_1\quad ,\quad \tilde  \beta '= d_-k_2\quad
,\quad \tilde  \rho '={1\over 2} d_-^2 \cr}\eqno(2.21)
$$

Notice that the gravitational part of the amplitude (integrals over
$z_i$) is
the same as in the case  without screening charges. The
integrals over $t_i$ and $r_j$ (matter contributions) have been also
calculated by Dotsenko and Fatteev\ref{12} (See their formula (B.10));
the result turns out to be
$$\eqalign{
{\cal A}_3^{nm}&=\! \left( {\mu \over \pi}\right)^s\Gamma(-s)
\Gamma(s\! +\! 1)
\pi^{s+n+m}\tilde \rho^{-4nm}
[\Delta(1\! -\! \tilde \rho)]^n[\Delta (1 \! -\!\tilde \rho']^m
\prod_{i=1}^m\Delta(i\tilde \rho'\! -\! n)\prod_{i=1}^n
\Delta(i\tilde \rho)\cr
&\times \prod_{i=0}^{m-1}
\Delta(1-n+\tilde \alpha' +i\tilde \rho')\Delta (1-n+\tilde \beta'
+i\tilde\rho' )
\Delta (-1+n-\tilde \alpha' -\tilde \beta' -(n-1+i)\tilde \rho')\cr
&\times \prod_{i=0}^{n-1}
\Delta(1+\tilde \alpha +i\tilde \rho)\Delta (1+\tilde \beta
+i\tilde\rho )
\Delta (-1+2m-\tilde \alpha -\tilde \beta -(n-1+i)\tilde \rho)\cr
&\times [\Delta (1-\rho)]^s\prod_{i=1}^s \Delta(i\rho)\prod_{i=0}^{s-1}
\Delta(1+ \alpha +i \rho)\Delta (1+\beta +i\rho )
\Delta (-\! 1 \! -\! \alpha\! -\! \beta\! -\! (s\! -\! 1\! +\! i)
\rho)\cr}\eqno(2.22)
$$

Using the same kinematics $(k_1\, ,\, k_3\ge \alpha_0\, ,\, k_2 < \alpha_0)$
we eliminate $\beta$ again using (2.5), (2.9) and
momentum conservation $\sum
\limits_{i=1}^3k_i\!+\!nd_+\!+\!md_-\!=\!2\alpha_0$,
$$
\beta=\cases{\rho(m+1-s)+n\quad ,\quad \alpha_0>0\cr
\-1-m-(s+n)\rho \quad .\quad \alpha_0<0\cr }
$$
It is easy to see, assuming $s\ge m+2$, that for $\alpha_0>0$ the
amplitude
vanishes again due to a factor $\Gamma(-n)$ in the denominator in the
gravitational part of the amplitude. Therefore we concentrate now on the
$\alpha_0<0$ case where we have
$$
\eqalign{
\tilde \alpha&= \alpha -2\rho \quad ,\quad \tilde \alpha'=-2+\tilde
\rho \alpha
\cr
\tilde \beta &= m-1+(s+n)\rho \quad ,\quad \tilde \beta ' = s+n+\rho
^{-1}(m-1)\cr
\tilde \rho &= -\rho \quad \quad ,\quad \tilde \rho ' =- \rho^{-1}\cr
}\eqno(2.23)
$$
Substituting in (2.22) we obtain:
$$
{\cal A}_3^{nm}=\left( {\mu \over \pi}\right) ^s\Gamma(-s)\Gamma(s+1)
\pi^{s+n+m}(\tilde \rho)^{-4nm}(C_{M+G}D_ME_{M+G})
\eqno(2.24)
$$
where
$$\eqalignno{C_{M+G}=& \left[ \Delta(1+\rho^{-1})\right]^m
\left[ \Delta(1+\rho)\right]^n
\prod_{i=1}^m\Delta(i\rho^{-1}-n)\prod_{i=1}^n\Delta
(-i\rho )\cr
\times &\left[ \Delta (1-\rho)\right] ^s \prod_{i=1}^s\Delta(i\rho)\cr
\times &
\prod_{i=0}^{n-1}\Delta(m+(s+n-i)\rho)\prod_{i=0}^{s-1}\Delta(-m-
(s+n-i)\rho)\cr
\times &\prod_{i=0}^{m-1}\Delta(1+s+(m-1-i)\rho^{-1}) &(2.25a)\cr
D_M=&\prod_{i=0}^{m-1}\Delta(-1-n+\rho^{-1}\alpha-i\rho^{-1})\Delta
(1-s-\rho^{-1}\alpha +i\rho^{-1})&(2.25b)\cr
E_{M+G}=&\prod_{i=0}^{n-1}\Delta(1+\alpha-(i+2)\rho)
\Delta(m-\alpha - (s-1-i)\rho)\cr
&\prod_{i=0}^{s-1}\Delta(1+\alpha+i\rho)\Delta(m-\alpha
+ (n+1-i)\rho)&(2.25c)\cr}
$$
To get a simple expression for ${\cal A}_3^{nm}$ we look for
$\Delta(\rho\!-\!\alpha) \Delta (\rho (s\!-\!n\!+\!1)\!+\!\alpha \!-\!m\!+\!1)
\Delta(-m\rho^{-1}\!-\!(s\!+\!n))$ which corresponds to $\prod_{i=1}^3
\Delta({1\over 2}
(\beta_i^2-k_i^2)) $. We expect that these terms show up in
the result. For
example:
$$
E_{M+G}=\! \prod_{i=2}^{n+1}\!\Delta(1\!+\!\alpha\!-\!i\rho)
\prod_{i=0}^{s-1}\Delta(1\!+\!\alpha\!+\!i\rho)\!\!
\prod_{i=s-n}^{s-1}\! \! \Delta(m\!-\!\alpha\! -\! i\rho)\! \!
\prod_{i=-(n+1)}^{s-n-2}\! \!\Delta(m\!-\!\alpha\!-\!i\rho)
\eqno(2.25d)
$$
Using $\Delta (x)\Delta (1-x)=1$, and $\quad \Delta (x+1)=-x^2
\Delta (x)$, we easily get:
$$
\eqalign{
E_{M+G}&=(-)^{m(s+n+1)}\Delta(1-m+\alpha+(s-n+1)\rho)
\Delta(\rho-\alpha)\cr
&\times\prod_{1-s}^{n+1}(m-1-\alpha+i\rho)^2 (m-2-\alpha+i\rho)
\cdots
(-\alpha+i\rho)^2\cr}
\eqno(2.26a)
$$
Analogously, we also arrive at
$$
\eqalignno{
D_M&=(-)^{m(n+s+1)}\rho^{2m(n+s+1)}\prod_{1-s}^{n+1}
\left[{\Gamma(-\alpha+i\rho )\over \Gamma (m-\alpha+i\rho)}\right]^2
&(2.26b)\cr
C_{M+G}&={(-)^s\rho^{-2(s+n)+2m(n-s)}\over \Gamma (-s)\Gamma(s+1)}
\left[ \Delta(1+\rho ^{-1})\right]^m \left[\Delta(1+\rho )\right]^n
\left[\Delta(1-\rho )\right]^s &(2.26c)\cr
}
$$
 Now substituting (2.26) into (2.24) we have
$$
{\cal A}_3^{nm}=\left[ \mu \Delta (-\rho)\right]^s\left[ -\pi \Delta
(\rho^{-1})\right]^m\left[ -\pi \Delta(\rho)\right]^n
\prod_{i=1}^3(-\pi)\Delta\left( {1\over2}(\beta_i^2-k_i^2)\right)
\quad .\eqno(2.27)
$$
Therefore redefining the screening operators as
$$\eqalignno{
e^{id_+X}&\to {1\over \Delta (\rho)}e^{id_+X} &(2.29a)\cr
e^{id_-X}&\to {1\over \Delta (\rho^{-1})}e^{id_-X}\quad ,&(2.29b)\cr}
$$
the operators $T_{k_i}$ and the cosmological constant $\mu $ as before
(see (2.17)), we get the very simple result:
$$
{\cal A}_3^{nm}=\mu ^s\quad ,\eqno(2.29)
$$
which should be compared to (2.18).
This result has been also obtained by Di Francesco and Kutasov\ref{6,16}
Note that
the factors $\Delta(\rho)$ and $\Delta(\rho^{-1})$ can be easily
understood; the
screening operators are renormalized like the tachyon vertex operators $T_k$
with vanishing dressing $\beta(k)$.

\vskip 1cm
\penalty-200

\noindent {\bf 2.3- $N$-Point tachyon amplitude ($N\ge 4$) with an
arbitrary
number of  screening charges}
\vskip .5cm
\nobreak
Repeating the zero-mode trick in the most general case of an $N$-point
function with arbitrary screening charges we
have\newfoot{$^{3)}$}{We have
absorbed a factor $\pi^3/\alpha_+$ in the measure of the path
integral.}:
$$
\eqalign{
{\cal A}_N^{nm}&=(-\pi)^3\left({\mu\over \pi}\right)^s\Gamma(-s)
\prod_{i=1}^N\int d^2z_i\prod_{j=1}^n\int {d^2t_j\over n!}\prod_{k=1}^m
\int {d^2r_k\over m!}\cr
&\times\prod_{l=1}^s\int d^2w_l\left\langle e^{ik_iX(z_i)}
e^{id_+X(t_j)}
e^{id_-X(r_n)}\right\rangle_0\left\langle e^{\beta_i\phi(z_i)}
e^{\alpha_+
\phi(w_l)}\right\rangle_0\quad ,\cr}
\eqno(2.30)
$$
where $s=-{1\over \alpha_+}(\sum_{i=1}^N\beta_i+Q)$. Fixing the $SL(2,
\Complex)$ symmetry we get:
$$
\eqalign{
{\cal A}_N^{nm}&=(-\pi)^3\left( {\mu\over \pi}\right)^s \Gamma(-s)
I_N^{nm}
\quad ,\cr
I_N^{nm}&=\int \prod_{j=4}^Nd^2z_j\vert z_j\vert^{2\alpha'_j}
\vert 1-z_j\vert
^{2\beta_j}\prod_{i<j=4}^N\vert z_i-z_j\vert ^{4\rho'_{ij}}\cr
&\times\int \prod_{i=1}^sd^2w_i\vert w_i\vert ^{2\alpha}\vert 1-w_i
\vert ^{2\beta}\prod_{i<j=1}^s\vert w_i-w_j\vert ^{4\rho}\prod_{i=1}^s
\prod_{j=4}^N\vert w_i-z_j\vert ^{2p_j}\cr
&\times \prod_{i=1}^nd^2t_i\vert t_i\vert ^{2\tilde \alpha}
\vert 1-t_i\vert
^{2\tilde \beta}\prod_{i<j}^n\vert t_i-t_j\vert ^{4\tilde \rho}
\prod_{i=1}
^n\prod_{j=4}^N\vert z_j-z_i\vert ^{2\tilde \alpha_j}\cr
&\times\int \prod_{i=1}^md^2r_i\vert r_i\vert ^{2\tilde\alpha'}
\vert 1-r_i
\vert^{2\tilde\beta'}\prod_{i<j=1}^m\vert r_i-r_j\vert ^{4\tilde\rho'}
\prod_{i=1}^m\prod_{j=4}^N\vert z_j-r_i\vert ^{2\tilde\alpha'_j}\cr
&\times \prod_{i=1}^n\prod_{j=1}^m\vert t_i-r_j\vert ^{-4}\quad ,\cr}
\eqno(2.31)
$$
where $\alpha,\beta,\rho,\tilde\alpha,\tilde\beta,\tilde\rho,
\tilde\alpha',
\tilde\beta',\tilde\rho'$ are defined as
$$
\eqalign{
\alpha_j&=k_1k_j-\beta_1\beta_j\quad ,\quad \tilde \alpha_j=d_+k_j\cr
\beta_j&=k_2k_j-\beta_2\beta_j\quad ,\quad \tilde\alpha_j=d_-k_j\cr
\rho_{lj}&={1\over 2}(k_lk_j-\beta_l\beta_j)\quad ,\quad 4\le j,l\le
N\quad .\cr}
\eqno(2.32)
$$

The integral $I_N^{nm}$ for the case $n=m=0$ has been calculated by
Di Francesco
and Kutasov\ref{6}. We shall use the same technique for arbitrary
$n,m$. First
we notice that translation invariance $(w_i\to 1-w_i\, ,\,
z_i\to 1-z_i\, ,\,
t_i\to 1-t_i \, ,\, r_i\to 1-r_i)$ implies $(\alpha \leftrightarrow \beta\, ,\,
\alpha'_j
\leftrightarrow \beta'_j\, ,\, \tilde\alpha\leftrightarrow \tilde\beta\, ,\,
\tilde\alpha'\leftrightarrow \tilde
\beta')$ so after the elimination of the remaining
parameters
as a function of $\alpha,\beta,p_j$ and $\rho$ $(j=4,5,\cdots, N-1)$,
$I_N^{nm}$
exhibits an $\alpha$-$\beta$ symmetry
$$
I_N^{nm}(\alpha,\beta,p_j,\rho)=I_N^{nm}(\beta,\alpha,p_j,\rho)
\quad .\eqno(2.33)
$$
Similarly by the inversion of all variables $w_i, z_i, t_i, r_i$ we have:
$$
I_N^{nm}(\alpha,\beta,p_j,\rho)=I_N^{nm}(-2-\alpha-\beta-2\rho(s-1)
-p_N-P,
\beta,p_j,\rho)
\eqno(2.34)
$$
where $P=\sum_{j=4}^{N-1}p_j$. Further information about $I_N^{nm}$
can be
obtained in the limit $\alpha\to \infty $ (or $\beta \to \infty$),
by using a
technique applied by Dotsenko and Fatteev\ref{12} in the case of
contour
integrals. Take for instance the simple case:
$$
I(\alpha,\beta)=\int d^2w\vert w\vert ^{2\alpha}
\vert 1-w\vert ^{2\beta}
=\pi \Delta (1+\alpha)\Delta(1+\beta)\Delta (-1-\alpha-\beta)
\eqno(2.35)
$$
by making a change of variables $w\to e^{-{\tilde w\over \alpha}}\,
\left(
w^* \to e^{-{\tilde w^*\over \alpha}}\right)$ we have :
$$
I(\alpha\to \infty,\beta)\approx \alpha^{-2-2\beta}
\tilde I(\beta)\quad
.\eqno(2.36)$$

This large-$\alpha$ behaviour can be checked by using Stirling's
formula ($\Gamma(\alpha\!+\!B)\sim \alpha^B\Gamma(\!\alpha\!)$) on the r.h.s.
of
equation (2.35).
Applying this technique to $I_N^{nm}$ we get:
$$
I_N^{nm}\approx \alpha^{2\beta +2\rho(s-N-n+3)+2P-2m}\eqno(2.37)
$$
where we have used the kinematics:
$$
k_1,k_2,\cdots ,k_{N-1}\ge \alpha_0\, ,\, k_N<\alpha_0 \eqno(2.38)
$$
and assumed $\alpha_0<0$, which permits us to eliminate the remaining
parameters in terms of $\alpha,\beta,p_j$ and $\rho$ as follows:
$$
\eqalign{
p_N&=-1-m-\rho(N+s+n-3)\cr
\beta'_j&=\beta+p_j-2\rho\cr
\beta'_N&=m-1+(\rho-\beta)(N+s+n-3)-m\rho^{-1}\beta\cr
\rho'_{jl}&={1\over 2}(p_j+p_l)-\rho \cr
\rho'_{jN}&={m-1\over 2}+{(\rho-p_j)\over 2}(N+s+n-3)
-{m\rho^{-1}\over 2}p_j
\quad , \cr
\tilde \alpha&=\alpha-2\rho\quad ,\quad \tilde \alpha'=
\alpha\rho^{-1}-2\quad
,\quad \tilde\rho=-\rho \quad ,\cr
\tilde \beta &=\beta-2\rho\quad ,\quad \tilde \beta'=\beta
\rho^{-1}-2 \quad
,\quad \tilde \rho'=-\rho^{-1}\cr
\tilde\alpha_j&=p_j-2\rho\quad ,\quad \tilde \alpha'_j=
\rho^{-1}p_j-2\quad ,\cr
\tilde \alpha_N&=m-1+\rho(N+s+n-3)\quad ,\quad \tilde \alpha'_N=
N+s+n-3+
\rho^{-1}(m-1)\quad .\cr}\eqno(2.39)
$$
where $4\le j,l \le N-1 $.
Notice that eliminating $p_N$ the symmetry under inversion implies:
$$
I_N^{nm}(\alpha,\beta,p_j,\rho)=I_N^{nm}(m-1-P-\alpha-\beta+\rho(N+n
-1-s),
\beta,p_j,\rho)\eqno(2.40)
$$
It is not difficult to check (using Stirling's formula) that
the following
Ansatz is consistent with (2.33), (2.37) and (2.40):
$$
\eqalignno{
{\cal A}_N^{nm}&=f_N^{nm}(\rho,p_j)\Delta(\rho\!-\!\alpha)
\Delta(\rho\!-\!\beta)
\Delta(1\!-\! m\!+\!P\!+\!\alpha\!+\!\beta\!+\!\rho(s\!+\!2\!
-\!N\!-\!n)) &
(2.41a)\cr
{\cal A}_N^{nm}&=f_N^{nm}(\rho,p_j)\prod_{j=1}^3\Delta\left(
{1\over 2}(\beta_j^2-k_j^2)\right) & (2.41b)\cr}
$$

Following Di Francesco and Kutasov\ref{6}, we can fix $f_N^{nm}(\rho,p_j)$
by using
the 3-point function ${\cal A}_3^{nm}$
through\newfoot{$^{4)}$}{Notice that
$\lim \limits _{k_j\to 0}\beta_j=\alpha_+$, thus $s=
-{1\over \alpha_+}\! \!
\left( \sum \limits _{j=1}^N\beta_j\! +\! Q\right) \! \to
\tilde s+3-N$ where
$\tilde s={-1\over \alpha_+}\! \! \left(
\sum \limits _{j=1,2,N}\beta_j\! +\! Q
\right)$.}:
$$
{\cal A}_N^{nm}(k_1,k_2,k_j\to 0,k_N)=(-\pi)^{N-3}{\partial\over
\partial \mu} ^{N-3}{\cal A}_3^{nm}(k_1,k_2,k_N)\quad ,\quad
3\le j\le N-1
\eqno(2.42)$$

Now using the result for ${\cal A}_3^{nm}$ (formula (2.27)) we get:
$$
f_N^{nm}(\rho,p_j)\! =\! [-\pi\Delta(\rho^{-1})]^m [-\pi\Delta(\rho)]^n
\! \! \left( \! {\partial ^{N-3}\over \partial _\mu}
\mu^{s+N-3}\!\right) \! \!
[\Delta(-\rho)]^s\! \! \prod_{j=4}^N(-\pi)\Delta({1\over 2}(\beta_j^2
\! -\! k_j^2))\quad .\eqno(2.43)
$$
Back in (2.41) we have
$$
\displaylines{
{\cal A}_N^{nm}=(s+N-3)(s+N-4)\cdots (s+1)\left[\mu\Delta (-\rho )\right]^s
\hfill\cr
\hfill \left[ -\pi\Delta (\rho^{-1})\right]^m
\left[ -\pi\Delta(\rho )\right]^n
\prod_{j=1}^N (-\pi )\Delta ({1\over 2} (\beta_j^2-k_j^2))
 \quad ,\quad(2.44)\cr}
$$
therefore, redefining the screening operators, $T_{k_j}$ and $\mu$ as
before, we have:
$$
{\cal A}_N^{nm}={\partial^{N-3}\over\partial\mu}\mu^{s+N-3}\eqno(2.45)
$$
which is a remarkable result. We generalize this technique
to the NS sector of the supersymmetric theory in the next section.

\vskip 1cm
\penalty-200
\centerline {\bf 3- Supersymmetric Correlators }
\vskip .5cm
\noindent {\bf 3.1- The 3-point  fermionic NS correlator}
\vskip .5cm
\nobreak

In a recent paper\ref{15} we have calculated the 3- and 4-point NS
correlations
functions using DHK formulation\ref{13} of super Liouville theory
coupled to
superconformal matter on the sphere without screening charge.
The total action $S$
is given by $S=S_{SL} +S_M $ where
$$
\eqalign{
S_{SL}=&{1\over 4\pi}\int d^2{\bf z}\hat E\left( {1\over 2}
\hat D_\alpha
\Phi_{SL}\hat D^\alpha \Phi_{SL}-Q\hat Y\Phi_{SL}-4i\mu
e^{\alpha_+\Phi_{SL}}\right)\quad ,\cr
S_M =&{1\over 4\pi}\int d^2{\bf z}\hat E({1\over 2}\hat D_\alpha
\Phi_{M}\hat
 D^\alpha \Phi_{M} + 2i\alpha_0\hat Y\Phi_M)\quad ,\cr}\eqno(3.1)
$$
where $\Phi_{SL}, \Phi_M$ are super Liouville and matter superfields
respectively. The matter sector has the central charge $\hat c
_m=1-8\alpha_0^2$. Analogous to the bosonic case the parameters $Q$ and
$\alpha_\pm $ are given by (compare with (2.2))
$$Q=2\sqrt {1+\alpha_0^2}\quad ,\quad \alpha_\pm =-{Q\over 2} \pm
{1\over
2}\sqrt {Q^2-4}=-{Q\over 2}\pm \vert \alpha_0 \vert\quad ,\quad \alpha
_+\alpha _-=1 \quad .\eqno(3.2)$$
The (gravitationally dressed) primary superfields $\tilde \Psi _{NS}$ are
 given by
$$
\tilde  \Psi _{NS}({\bf z}_i,k_i)=d^2{\bf z}\hat E
e^{ik\Phi_M({\bf z})}e^{\beta(k)\Phi_{SL}({\bf z})}$$
$$ {\rm where } \quad \quad \beta (k)= -{Q\over 2} +\vert k-\alpha _0 \vert
\quad .\eqno(3.3)
$$

In what follows we review the calculation of the  three-point function of
the primary
superfield $\tilde \Psi _{NS} $, that is:
$$
\left\langle \prod_{i=1}^3\int \tilde  \Psi _{NS}({\bf
z}_i,k_i)\right\rangle \equiv \int [{\cal D}_{\hat E}\Phi_{SL}]
[{\cal D}_{\hat E} \Phi_M]
\prod _{i=1}^3\tilde  \Psi _{NS}({\bf z}_i,k_i)e^{-S}
\quad .\eqno(3.4)
$$

The method will closely parallel the bosonic case.
After integrating over the bosonic zero modes we get
$$
\eqalign{&\left\langle \prod_{i=1}^3\int \tilde  \Psi _{NS}({\bf
z}_i,k_i)\right\rangle \equiv 2\pi \delta \left( \sum
_{i=1}^3k_i-2\alpha_0\right) {\cal A}(k_1,k_2,k_3)\quad ,\cr
&{\cal A}(k_1,k_2,k_3) = \Gamma (-s)({-\pi \over 2})^3({i\mu \over \pi
})^s\left\langle \!\int \! \prod _{i=1}^3d^2\tilde  {\bf
z}_ie^{ik_i\Phi_M(\tilde  {\bf z}_i)}e^{\beta_i\Phi_{SL}(\tilde  {\bf
z}_i)}\left(\! \int \! d^2{\bf z}e^{\alpha_+\Phi_{SL}({\bf z})}
\! \right)
^s\right\rangle_0  \cr }\eqno(3.5)
$$
where $\langle \cdots \rangle _0$ denotes again the expectation value
evaluated in
the free theory $(\mu =0)$  and we have absorbed the factor
$[\alpha _+(-\pi/2)^3]^{-1}$ into the normalization
of the path integral. the parameter $s$ is
defined as in the bosonic case (see (2.9)).

For $s$ non-negative integer, we have
$$
\eqalign{&{\cal A}(k_1,k_2,k_3) = \Gamma (-s)({-\pi \over 2})^3
({i\mu \over
\pi})^s\cr
&\times \int \prod _{i=1}^3d^2\tilde  {\bf z}_i\prod_{i=1}^sd^2{\bf
z}_i\prod _{i<j}^3\vert \tilde  {\bf z}_ij\vert
^{2k_ik_j-2\beta_i\beta_j}\prod_{i=1}^3\prod _{j=1}^s\vert \tilde
z_i-z_j-\tilde  \theta_i\theta_j\vert ^{-2\alpha_i\beta_i}
\prod_{i<j}^s\vert
{\bf z}_{ij}\vert ^{-\alpha_+^2}\cr
&=\Gamma (-s)({-\pi \over 2})^3({i\mu \over \pi})^s\int \prod
_{i=1}^sd^2z_id^2\tilde  \theta \prod_{i=1}^s\vert z_i+\tilde  \theta
\theta_i\vert ^{-2\alpha_+\beta_1}\prod _{i=1}^s\vert 1-z_i\vert
^{-2\alpha_+\beta_2}
\prod_{i<j}^s\vert {\bf z}_{ij}\vert^{-2\alpha_+^2}\cr
}\eqno(3.6)
$$

The $\widehat {SL_2}$ volume is treated analogously to the bosonic
case. Indeed,
the generators of the superconformal transformations on the
$(z,\theta )$
variables
$$
\eqalign{L_0&=z\partial _z+{1\over 2}\theta\partial _\theta -j\cr
L_1&=\partial _z \cr
L_{-1}&=z^2\partial _z + z\theta\partial_\theta -2jz \cr
Q^{1/2} &=i\sqrt{{1\over 2}}(\partial_\theta + \theta\partial_z)\cr
Q^{-1/2} &=i\sqrt{{1\over 2}}(z\partial_\theta + z\theta\partial _z-2j\theta)
\cr}\eqno(3.7)
$$
imply that we can fix $z_1=0\, ,\, z_2=1\, ,\, z_3=\infty\, ,\,
\theta_2=\theta_3=0\, ,\, \theta_1=\theta$. The integral is the
supersymmetric
generalization of (B.9) of Ref.[12]. Alternatively, it is
expressed in  components
($\Phi_{SL}=\phi +\theta\psi +\bar \theta \bar \psi$):
$$
\eqalign{&{\cal A}(k_1,k_2,k_3) = \Gamma (-s)({-\pi \over 2})^3
({i\alpha_+^2\mu
\over \pi})^s\beta_1^2\cr
&\int \prod _{i=1}^sd^2z_i
\prod_{i=1}^s\vert z_i\vert ^{-2\alpha_+\beta_1}\vert
1\! -\!  z_i\vert ^{-2\alpha_+\beta_2}\prod_{i<j}^s\vert z_i\! -\! z_j
\vert ^{-2\alpha_+^2}\langle \overline \psi \psi (0)\overline \psi
\psi(z_1)\cdots \overline \psi \psi (z_s)\rangle_0\quad .}\eqno(3.8)
$$
We first observe that this is non-vanishing only for
$s$ odd ($s=2l+1$). One
may
evaluate $\langle \overline \psi \cdots \overline \psi \rangle_0$
and $\langle
\psi \cdots \psi\rangle_0$ independently. Since the rest of the
integrand is
symmetric, one may write the result in a simple form by relabelling
coordinates: (compare with $A_3$ bosonic formula (2.12)).
$$
\eqalign{{\cal A}(k_1,k_2,k_3)& = \Gamma (-s)({-\pi \over 2})^3{1\over
\alpha_+^2} ({i\alpha_+^2\mu \over \pi})^s
\alpha^2(-1)^{{s+1\over 2}}s!!\cr
\times &\int \prod _{i=1}^sd^2z_i\prod_{i=1}^s
\vert z_i\vert ^{2\alpha}\vert
1-z_i\vert ^{2\beta} \prod_{i<j}^s\vert z_i-z_j\vert^{4\rho}
\prod_{i=1}^{s-1\over 2}\vert z_{2i-1}-z_{2i}\vert ^{-2}
\vert z_s\vert^{-2}\quad . \cr
{\rm With  }\, &\, z_s\equiv w\, ,\, z_{2i-1}
\equiv \zeta_i \, ,\, z_{2i} \equiv \eta_i\quad , {\rm we \, have: }\cr
{\cal A}(k_1,k_2,k _3)&=-i{-\pi\over 2}^3\Gamma(-s)\Gamma(s+1){1\over \alpha_
+^2}\left(
{\alpha_+^2\mu \over \pi }\right) I^l(\alpha,\beta;\rho)\cr}\quad ,\eqno(3.9)
$$
where
$$
\eqalign{
&I^l(\alpha,\beta;\rho)\cr
&={1\over 2^mm!}\alpha^2\int d^2w\prod_{i=1}^md^2\zeta_id^2\eta_i
\vert w\vert^{2\alpha-2} \vert 1-w\vert ^{2\beta}
\prod_{i=1}^m\vert w-\zeta_i\vert ^{4\rho}
\vert w-\eta_i\vert ^{4\rho}\cr
&\times \prod_{i=1}^m\vert \zeta_i\vert ^{2\alpha}
\vert 1-\zeta_i\vert^{2\beta} \vert 1-\eta_i \vert ^{2\beta}
\prod_{i,j}^m\vert
\zeta_i-\eta_j^{4\rho}
\prod _{i<j}^m\vert \zeta_i-\zeta_j\vert ^{4\rho}\vert
\eta_i -\eta_j\vert ^{4\rho}
\prod_{i=1}^m\vert \zeta_i-\eta_i\vert ^{-2}\quad ,\cr
}\eqno(3.10)
$$
and $\alpha=-\alpha_+\beta_1\, ,\,
\beta=-\alpha_+\beta_2\, ,\, \rho =-{1\over
2} \alpha_+^2$.
In the Ref.[15] we calculated $I^l$ in detail by using the symmetries
$I^l(\alpha,\beta;\rho)=I^l(\beta,\alpha;\rho)$, $
I^l(\alpha,\beta;\rho)=I^l(-1-\alpha-\beta-4l\rho,\beta;\rho)$ and looking at
its large $\alpha$ behavior (see Ref.[15]) we obtained:
$$
\eqalign{I^l(\alpha,\beta;\rho)=&-{\pi^{2l+1}\over 2^{2l}}
\left[ \Delta \left({1\over 2}-\rho \right) \right]^{2l+1}
\prod_{i=1}^l\Delta (2+\rho)
\prod_{i=1}^{l}\Delta  \left( {1\over 2}+(2i+1) \rho \right) \cr
&\times\prod_{i=0}^l\Delta( 1+\alpha +2i\rho) \Delta( 1+\beta+2i
\rho) \Delta( -\alpha-\beta+(2i-4l)\rho)\cr
&\times\prod_{i=1}^l\Delta( {1\over 2}+\alpha+(2i-1
)\rho) \Delta( {1\over 2}+\!\beta\!+\!(2i\!-\!1)
\rho) \Delta( -{1\over 2}\!-\!\alpha\!-\!\beta\!+\!(2i\! -\!4l\!-\!1)\rho)
\cr}\eqno(3.11)
$$

 We can choose, without loss of generality $k_1,k_3\ge\alpha_0, k_2\le
\alpha_0$. We proceed now as in the bosonic case and we have
(compare with (2.14))
$$
\beta=\cases {\rho (1-s)\; \; (\alpha_0>0)\cr
-{1\over 2}-\rho s \; \; (\alpha_0<0)\quad .\cr}\eqno(3.12)
$$

Now we are ready to write down the amplitude. For $\alpha_0<0$ we have the
non-trivial amplitude:
$$\eqalignno{
{\cal A}(k_1,k_2,k_3)&= ({-i\pi\over 2})^3
\left[ {\mu\over 2}\Delta\left({1\over 2}-\rho\right) \right]^s
\Delta\left( {1\over 2}-{s\over 2}
\right)\Delta \left( 1+\alpha-(s-1)\rho\right) \Delta\left(
{1\over 2}-\alpha+\rho\right)\cr
&=\left[ {\mu\over 2}\Delta\left( {1\over 2}-\rho\right)\right]^s
\prod_{j=1}^3(-{i\pi\over 2})\Delta\left( {1\over 2}
[1+\beta_j^2-k_j^2]\right) &(3.13)\cr }
$$

 By redefining the cosmological constant and the primary superfield
$\tilde \Psi_{NS}$
$$
\mu  \to  {2\over \Delta \left( {1\over 2}-\rho\right) }\mu
\quad,\quad
\tilde \Psi_{NS}(k_j)\to {1\over (-{i\over 2}\pi)\Delta\left( {1\over
2}[1+\beta_j^2-k_j^2]\right)}\tilde \Psi_{NS}(k_j)\quad , \eqno(3.14)
$$
we get
$$
{\cal A}(k_1,k_2,k_3)=\mu ^s\quad . \eqno(3.15)
$$
As in the bosonic case we have a remarkably simple result. The only
differences with respect to the bosonic case are in  the details of the
renormalization factors.
Compare (3.14) with (2.17). Note that the singular point at
the renormalization
of the cosmological constant is $\rho=-1$ in the bosonic case, which
corresponds to $c=1$, and $\rho=-{1\over 2}$ in the
supersymmetric case,
corresponding to $\hat c=1$ or $c=3/2$, as it should.
\vskip 1cm
\penalty-200
\noindent {\bf 3.2- The 3-point NS correlator with arbitrary s.c.
$({\cal A}_3^{nm})$}
\vskip .5cm
\nobreak

Now we shall generalize the above result to the case which includes
screening
charges in the supermatter sector. We consider $n$
charges $e^{id_+\Phi_M}$
and $m$ charges $e^{id_-\Phi_M}$, where $d_\pm $ are solutions of
the equation
 ${1\over 2}d(d-2\alpha_0)={1\over 2}$. After integrating over the
matter and
Liouville zero modes we get
$$
\eqalignno{
&\left\langle\prod_{i=1}^3\int \tilde  \Psi_{NS}(\tilde  {\bf z}_i,k_i)
\prod_{i=1}^n\int {d^2{\bf t}_i\over n!}e^{id_+\Phi_M({\bf t}_i)}
\prod_{i=1}^m\int
{d^2{\bf r}_i\over m!}e^{id_-\Phi_M({\bf r}_i)}  \right \rangle \cr
&\equiv 2\pi \delta \left( \sum _{i=1}^3k_i+nd_+
+md_- -2\alpha_0\right)
{\cal A}_3^{nm}(k_1,k_2,k_3)\cr
&{\cal A}_3^{nm}(k_1,k_2,k_3)=\Gamma(-s)({-\pi\over 2})^3({i\mu
\over \pi})^s\! \left\langle
\prod_{i=1}^n\int {d^2{\bf t}_i\over n!}e^{id_+\Phi_M({\bf
t}_i)} \prod_{i=1}^m\int {d^2{\bf r}_i\over m!}e^{id_-\Phi_M({\bf r}_i)}
\right.\cr
&\left.\times \int \prod_{i=1}^3d^2{\bf \tilde z}_i
e^{ik_i\Phi_M( \tilde {\bf
z}_i)}e^{\beta_i\Phi_{SL}({\bf \tilde z}_i)}
\left(\! \int d^2{\bf z} e^{\alpha_+
\Phi_{SL} ({\bf z})}\right)^s\!\right\rangle_0 \! ,&(3.16)\cr}
$$

Integrating over the Grasmann variables and fixing the
$\widehat {SL(2)}$ symmetry
as before $(\tilde  z_1=0\, ,\, \tilde  z_2=\infty\, ,\, \tilde
\theta_1=\theta\, ,\, \tilde  \theta_2=\tilde  \theta_3=0)$ we obtain
(using $d_+d_-=-\alpha_+\alpha_-=-1$)
$$\eqalignno{
{\cal A}_3^{nm}(k_1,k_2,k_3)
&=\Gamma(-s)\left( {-\pi\over 2}\right) ^3\left( {i\mu \alpha_+^2\over
\pi}\right) ^s{(-d_+^2)^n\over n!}{(-d_-^2)^m\over m!}\cr
&\times\prod _{i=1}^n\int d^2t_i\vert t_i\vert ^{-2d_+k_1}
\vert 1-t_i\vert^{-d_+k_2}\prod _{i<j}^n
\vert t_i-t_j\vert ^{2d_+^2}\cr
&\times \prod _{i=1}^m\int d^2r_i\vert r_i\vert^{-2d_-k_1}
\vert 1-r_i\vert^{-2d_-k_2}
\prod_ {i<j}^m\vert r_i-r_j\vert ^{2d_-^2}\prod_{i=1}^{n}
\prod_{j=1}^m\vert
t_i-r_j\vert^{-2}\cr
&\times \prod_{i=1}^s\int d^2z_i\vert z_i\vert ^{-2\alpha_+\beta_1}\vert
1-z_i\vert^{-2\alpha_+\beta_2}\prod_{i<j}^s
\vert z_i-z_j\vert ^{-2\alpha_+^2}\cr
&\times \left\langle (\beta_1^2\overline \psi\psi(0)-k_1^2\overline \xi
\xi(0))\prod _{i=1}^n\overline \xi \xi (t_i)
\prod _{i=1}^m\overline \xi \xi
(r_i) \prod_{i=1}^s\overline \psi \psi (z_i)\right\rangle_0 &(3.17)\cr }
$$

Since the vacuum expectation value of an odd number of
$\overline \psi \psi $
(or $\overline \xi \xi$ ) operators is zero we have only two
non-trivial cases:
in the first case $n+m=$ odd, $s=$ even and in the second one
$n+m=$ even, $s=$ odd. Thus we have
$$\eqalign{
{\cal A}_3^{nm}(k_1,k_2,k_3)
&=\Gamma(-s)\left( {-\pi\over 2}\right) ^3
\left( {i\mu \alpha_+^2\over
\pi}\right) ^s{(-d_+^2)^n\over n!}{(-d_-^2)^m\over m!}\cr
&\times \cases {I^{nm}_M(\tilde  \alpha,
\tilde \beta;\tilde \rho)\,\times \,
I^s_G(\alpha,\beta;\rho)\, ,\, n+m={\rm even }\, ,\,
s={\rm odd}\cr
{}\cr
J^{nm}_M(\tilde  \alpha, \tilde \beta;\tilde \rho)\,\times \,
J^s_G(\alpha,\beta;\rho)\, ,\, n+m={\rm odd }\, ,\, s={\rm
even}\cr}}\eqno(3.18)
$$
where
$$
\eqalignno{
I^{nm}_M(\tilde  \alpha,\tilde  \beta;\tilde  \rho)& =
\prod _{i=1}^n\int d^2t_i\vert t_i\vert ^{2\tilde  \alpha}\vert 1-
t_i\vert^{2\tilde  \beta}\prod _{i<j}^n\vert t_i-t_j\vert ^{2\tilde
\rho}\cr
&\times \prod _{i=1}^m\int d^2r_i\vert r_i\vert^{2\tilde \alpha'}
\vert 1-
r_i\vert^{2\tilde \beta'}\prod_ {i<j}^m\vert r_i-r_j\vert ^{2\tilde
\rho'}\prod_{i=1}^{n}\prod_{j=1}^m\vert t_i-r_j\vert^{-2}\cr
&\times \left\langle \prod _{i=1}^n\overline \xi \xi (t_i)
\prod_{i=1}^m\overline \xi \xi (r_i)
\right\rangle_0\quad , &(3.19)\cr}
$$
 $$
I^s_G(\alpha,\beta;\rho)=\alpha^2\int \prod _{i=1}^sd^2z_i
\prod_{i=1}^s\vert
z_i\vert ^{2\alpha}\vert 1-z_i\vert ^{2\beta}
\prod_{i<j}^s\vert z_i-z_j\vert^{4\rho}
\left\langle\overline \psi\psi (0)
\prod_{i=1}^s\overline \psi \psi
(z_i) \right\rangle _0\eqno(3.20a)
$$
with
$$\eqalign{
\tilde  \alpha &=-d_+k_1\quad ,\quad \tilde  \beta = -d_+k_2\quad ,\quad
\tilde  \rho =d_+^2 \cr
\tilde  \alpha '&=-d_-k_1\quad ,\quad
\tilde  \beta '= -d_-k_2\quad,\quad
\tilde  \rho '=d_-^2 \cr}\eqno(3.20b)
$$
and $\alpha ,\beta ,\rho $ are defined as before. Note that
$I^{nm}_M$ is the
supersymmetric generalization of (B.10) of Ref.[12]. The integral
$J^{nm}_M$
differs from $I^{nm}_M$ by the introduction of a factor
$\overline \xi \xi(0)$
and $J^s_G$ can be obtained from $I^s_G$  by dropping
$\overline \psi\psi(0)$.
Henceforth we assume,
for simplicity, $n+m=$ even, $s=$ odd. We will work out explicitly
only the
case $n,m$ even. However, the final result for the amplitude does
not depend on which case we choose. In the Appendix A we calculate
$I_M^{nm}$ for $n$ and $m$ even and we get:
$$
\eqalign{
&I^{nm}_M(\tilde  \alpha,\tilde \beta;\tilde \rho)=
(-)^{{n+m\over 2}}{\pi^{n+m}\over 2^{n+m}}n!m!
\left( -{\tilde  \rho\over2}\right) ^{-2nm}
\left[\Delta\left( {1\over 2}-{\tilde \rho\over 2}\right)\right] ^n
\left[\Delta\left( {1\over 2}-{\tilde \rho'\over 2}\right)\right] ^m
\cr
&\times \prod _1^{n\over 2}\Delta (i\tilde  \rho)\Delta\left( {1\over
2}+\tilde  \rho\left( i-{1\over 2}\right) \right) \prod _1^{m\over 2}
\Delta (i\tilde  \rho' -{n\over 2})\Delta\left( {1\over 2}-{n\over 2}
-\tilde  \rho'\left( i-{1\over 2}\right) \right) \cr
&\times \prod _{i=0}^{{n\over 2}-1}\Delta (1+\tilde \alpha + i\tilde
\rho )\Delta(1+\tilde  \beta +i\tilde  \rho)\Delta(m-\tilde  \alpha
-\tilde  \beta +\tilde  \rho (i-n+1))\cr
&\times \prod _{i=1}^{{n\over 2}}\Delta ({1\over 2}\! +\! \tilde \alpha
\!+\! (i \! - \! {1\over 2} )\tilde
\rho )\Delta({1\over 2}\! +\! \tilde  \beta \! + \! (i\! -\! {1\over 2})
\tilde  \rho)
\Delta(-{1\over 2}\! -\! \tilde  \alpha \! +\! m \! -\tilde  \beta \! +
\! \tilde  \rho (i\! -\! n\! +\! {1\over 2}))\cr
&\times \prod _{i=0}^{{m\over 2}-1}\Delta (1+\tilde \alpha' - {n\over 2}
+ i\tilde  \rho ')\Delta(1 - {n\over 2} + \tilde  \beta' +i\tilde
\rho' )\Delta( {n\over 2} -\tilde  \alpha' -\tilde  \beta '+\tilde
\rho' (i-m+1))\cr
&\times \prod _{i=1}^{{m\over 2}}\Delta ({1\over 2}\!
-\!  {n\over 2}\!  +\!
\tilde  \alpha '\! +\! (i\! -\! {1\over 2} )\tilde
\rho' )\Delta({1\over 2}\! -\! {n\over 2}\! +\! \tilde  \beta' \! +\!
(i\! - \! {1\over 2})\tilde  \rho')
\Delta(\! -{1\over 2}\! + {n\over 2}\! -\! \tilde  \alpha'\! -\! \tilde
\beta'\! +\! \tilde  \rho'(i\! -\! m \! + \! {1\over 2}))\cr}\eqno(3.21)
$$
In the case where $s=2l+1$ the gravitational contribution to
${\cal A}^{nm}(k_1,k_2,k_3) $, i.e, $I^s_G$ is just the same
as in the case
without screening charges, thus from the last section we have the
supersymmetric generalization of (B.9) of Ref.[12]:
$$
\eqalign{
I^s_G & = (-)^{{s-1\over 2}}{\pi^s\over 2^{s-1}}s!
\left[ \Delta \left(
{1\over 2} -\rho\right) \right] ^s
\prod_{i=1}^{{s-1\over 2}}\Delta (2i\rho)
\prod_{i=0}^{{s-1\over 2}}\Delta ( {1\over 2} + (2i+1)\rho) \cr
&\times \prod _{i=0}^{{s-1\over 2}}\Delta (1+\tilde \alpha + 2i\tilde
\rho )\Delta(1+\tilde  \beta +2i\tilde  \rho)\Delta(-\tilde  \alpha
-\tilde  \beta +2\tilde  \rho (i-s+1))\cr
&\times \prod _{i=0}^{{s-1\over 2}}\Delta ({1\over 2}+\tilde \alpha +
(2i-1)\tilde
\rho )\Delta({1\over 2} +\tilde  \beta +(2i - 1)\tilde  \rho)
\Delta(-{1\over 2}-\tilde  \alpha -\tilde  \beta +\tilde  \rho
(2i-2s+1))\cr}\eqno(3.22)
$$

To obtain ${\cal A}^{nm}(k_1,k_2,k_3) $ (see (3.18)) we have to
calculate
$I^{nm}\times I^s_G$. Using the same kinematics as
in the case without screening
charges,  it is easy to deduce:
$$
\beta=\cases {{n\over 2}+\rho(m-s+1)\quad \quad ,\alpha_0>0\cr
-{1\over 2}-{m\over 2}-(n+s)\rho\quad ,\alpha_0<0\cr }\eqno(3.23)
$$

As before, in the case $\alpha_0>0 $ the amplitude vanishes if we
assume\newfoot{$^{5)}$} {Actually the kinematics chosen here is
self only if $s>>n,m$.} $s\ge m+2$ because
there appears $\Gamma (-n)$ in the denominator of $I^s_G$ after the
substitution
of $\beta$ above  and consequently $I^s_G$ vanishes.
Therefore we
have to look at the $\alpha_0<0$ case to have a non-trivial
amplitude. In this
case we have (remember that $d_+=-\alpha_+\, ,\, d_-=\alpha_- \, ,
\, {\rm if}\, \alpha_0<0$):
$$\eqalign{
\tilde  \alpha&=\alpha-2\rho\quad ,\quad \tilde  \alpha'
=-1+{\rho^{-1}\alpha\over 2} \cr
\tilde  \rho&=-2\rho \quad \quad , \quad \tilde  \rho'= -{\rho^{-1}
\over
2} \cr
\tilde  \beta &=-\beta-1=(n+1)\rho +{m\over 2}-{1\over 2}\cr
\tilde  \beta'&={(n+s)\over 2}+{\rho^{-1}\over 4}(m-1)\quad .\cr}
\eqno(3.24)
$$
Substituting in (3.21) and (3.22) and using (3.18)
we have a very involved expression:
$$
\eqalign{
&{\cal A}_3^{nm}(k_1,k_2,k_3)\cr
&= \Gamma(-s)\left( {-\pi\over 2}\right) ^3\left( {i\mu
\over \pi}\right)^s\alpha_+^{2(s-1)}(2\rho)^{n-m}(-)^{n+m+s-1\over
2}{\pi^{s+n+m} \over 2^{m+n+s-1}}\rho^{-2mn}s! \cr
&\times \left[ \Delta({1\over 2}+\rho)\right]^n
\left[ \Delta({1\over
2}+{\rho^{-1}\over 4})\right]^m \cr
&\times \prod_{i=1}^{n\over 2}\Delta(-2i\rho)\Delta ({1\over 2}+\rho
(1-2i))\prod _{i=1}^{m\over 2} \Delta (-{n\over 2}-{i\rho^{-1}\over 2})
\Delta
\left( {1\over 2} -{n\over 2} +\left( {1\over 4}-{i\over 2}\right)
\rho^{-1}\right) \cr
&\times \prod _{i=0}^{{n\over 2}-1}\Delta \left( {1\over 2}
+{m\over 2}+\rho
(n+s-2i)\right) \prod_{i=1}^{n\over 2}\Delta\left( {m\over
2}+\rho(n+s+1-2i)\right) \cr
&\times \prod _{i=0}^{{m\over 2}-1}\Delta \left( 1+{s\over 2}
-{\rho^{-1}\over2} (i+{1\over 2}-{n\over 2})\right)
\prod_{i=1}^{m\over 2}\Delta\left( {1\over
2}+ {s\over 2}- {\rho^{-1}\over 2}(i-{m\over 2})\right) \cr
&\times \left[ \Delta ({1\over 2}-\rho)\right] ^s
\prod _{i=1}^{{s-1\over
2}}\Delta (2i\rho)\prod_{i=0}^{{s-1\over 2}}\Delta \left( {1\over
2}+(2i+1)\rho\right) \cr
&\times\prod_{i=0}^{{s-1\over 2}}\Delta ({1\over 2}-{m\over
2}+(2i-n-s)\rho)\prod _{i=1}^{{s-1\over 2}}\delta(-{m\over
2}+(2i-1-n-s)\rho)\cr
&\times \prod_{i=1}^{{n\over 2}}\Delta({1\over
2}+\alpha-\rho(2i+1))\Delta({1\over 2}+{m\over 2} -\alpha
-\rho(s-n-2+2i))\cr
&\times \prod_{i=0}^{{s-1\over 2}}\Delta({m\over 2}+{1\over 2}
-\alpha+(2i-s+n+2)\rho)
\prod_{i=1}^{{s-1\over 2}}\Delta({1\over 2}+\alpha+(2i-1)\rho)
\cr
&\times \prod_{i=1}^{{n\over 2}}\Delta({m\over
2}-\alpha-\rho(2i-n+s-1))\Delta(1+\alpha-2i\rho)\cr
&\times \prod_{i=0}^{{s-1\over 2}}\Delta(1+\alpha+
2i\rho)\prod_{i=1}^{{s-1\over 2}}\Delta({m\over 2}-\alpha
+(2i-s+n+1)\rho) \cr
&\times \prod_{i=1}^{{m\over 2}}\Delta({1\over
2}-{s\over 2}-{\rho^{-1}\over 2}(i+\alpha -{m\over 2}))
\Delta(-{n\over 2} -
{\rho^{-1}\over 2}(i-1-\alpha))\cr
&\times \prod_{i=1}^{{m\over 2}}\Delta(-{n\over 2}-{1\over 2}
-{\rho^{-1}
\over 2} (i-\alpha-{1\over 2}))\Delta(1-{s\over 2}-{\rho^{-1}\over
2}(i-{(m+1)\over 2}+\alpha)) \cr }\eqno(3.25)
$$

In order to obtain a simple expression for the amplitude we have
to combine in
each term  the matter and the gravitational parts as in the bosonic
case. The
calculation is more complicate now and we finally get
$$
\eqalignno{
{\cal A}_3^{nm}(k_1,k_2,k_3)&= \left( -{\pi\over 2}\right)^3
\left[ {\mu \over 2} \Delta ({1\over 2}-\rho)\right]^s
\left[-{i\pi\over 2}\Delta({1\over 2}+{\rho^{-1}\over 4})\right] ^m
\left[-{i\pi\over 2}\Delta({1\over 2}+\rho)\right] ^n\cr
&\times\Delta\left(\rho-\alpha+{1\over 2}\right)\Delta\left({1\over
2}-{n+s\over 2} -{m\rho^{-1}\over 4}\right) \Delta(1-{m\over 2}+\alpha
+(s-n-1)\rho)\cr
&=\left[ {\mu \over 2} \Delta \left( {1\over 2}-\rho\right) \right]^s
\left[-{i\pi\over 2}\Delta\left( {1\over 2}+{\rho^{-1}\over 4}\right)
\right] ^m
\left[-{i\pi\over 2}\Delta\left( {1\over 2}+\rho\right) \right] ^n
\cr
&\times\prod_{i=1}^3 \left( -{i\pi\over 2}\right) \Delta \left(
{1\over 2}
+{1\over 2}(\beta_i^2-k_i^2)\right)  &(3.26)\cr}
$$
Therefore after redefining the cosmological constant, the NS
operators and the screening charges
$$\eqalignno{
e^{id_+\Phi_M(t_i)}&\to
\left[ -{i\pi\over 2}\Delta\left( {1\over2}+\rho\right)
\right]^{-1}e^{id_+\Phi_M(t_i)} & (3.27a)\cr
e^{id_-\Phi_M(t_i)}&\to \left[ -{i\pi\over 2}\Delta\left( {1\over
2}+{\rho^{-1}\over 4} \right) \right]^{-1} e^{id_-\Phi_M(t_i)}
&(3.27b)\cr
\Psi_{NS}&\to \left[ -{i\pi\over 2} \Delta \left( {1\over 2}
+{1\over 2}(\beta_i^2-k_i^2)\right) \right]^{-1}\Psi_{NS}
& (3.27c)\cr
\mu &\to \left[ {1\over 2}\Delta \left( {1\over 2}-\rho\right)
\right]^{-1} \mu
& (3.27d)\cr }
$$
we obtain the very simple result:
$$
{\cal A}_3^{nm}(k_1,k_2,k_3)=\mu ^s\eqno(3.28)
$$
In view of the complexity of (3.25), the simplicity of the result is
remarkable.
\vskip 1cm
\penalty-200
\noindent  {\bf 3.3- $N$-point ($N\ge 4$) supersymmetric correlation
functions  with an arbitrary number  of screening charges}

\vskip 1cm
\nobreak

In this subsection we show that it is possible to obtain a simple
result for
the most general  case of a $N$-point amplitude with an arbitrary
number of
screening charges (${\cal A}^{nm}_N$). In that general case we have to
calculate the following integral\newfoot{$^{6)}$}{We are computing 1PI
amplitudes; see in this respect ref.[6]}
$$\eqalign{
{\cal A}^{nm}_N(k_1,&\cdots ,k_N)={\Gamma(-s)\over -\alpha_+}\left(
{i\mu \over
\pi}\right)^s\left\langle \prod_{i=1}^N\int d^2\tilde  {\bf
z}_ie^{ik_i\Phi_M(\tilde  {\bf z}_i)+\beta_i\Phi_{SL}(\tilde  {\bf
z}_i)}\right. \cr
&\left.\times \prod_{i=1}^n\int d^2{\bf t}_ie^{id_{}\Phi_M(\tilde
{\bf t}_i)}\prod_{j=1}^m\int d^2{\bf r}_je^{id_-\Phi_M
(\tilde  {\bf r}_j)}
\prod_{j=1}^s\int d^2{\bf z}_je^{i\alpha_+\Phi_{SL}(\tilde  {\bf z}_i)}
\right\rangle_0\cr }\eqno(4.1)$$
where $s=-{1\over \alpha_+}(\sum _{i=1}^N\beta_i+Q)$ and
$\sum_{i=1}^Nk_i+nd_++md_- =2\alpha_0$. After fixing the
$\widetilde  {SL_2}$
symmetry as before and integrating over the Grassmann variables
the amplitude
becomes
$$
\eqalignno{
{\cal A}^{nm}_N&=\Gamma(-s)\left( -{\pi\over 2}\right)^3\left(
{i\mu\alpha_+^2\over \pi} \right)^s(-d_+^2)^n(-d_-^2)^m\cr
&\times \prod_{j=4}^N\int d^2\tilde  z_j\prod_{i=1}^n\int d^2t_i
\prod_{i=1}^m\int d^2r_i\prod _{i=1}^sd^2w_i\vert w_i\vert ^{-2\alpha_+
\beta_1}\vert
1-w_i\vert ^{-2\alpha_+\beta_2}\cr
&\times \prod_{i<j}\vert w_i-w_j\vert^{-2\alpha_+^2}
\prod_{i=1}^s\prod_{j=4}^N\vert w_i-\tilde  z_j\vert^{-2\alpha_+
\beta_j} \cr
&\prod_{j=4}^N\vert \tilde  z_j\vert ^{2(k_1k_j-\beta_1\beta_j)}\vert
1-\tilde  z_j\vert ^{2(k_2k_j-\beta_2\beta_j)}\prod _{j<l=4}^N\vert
\tilde  z_j-\tilde  z_l\vert ^{2(k_jk_l-\beta_j\beta_l)}\cr
&\times\prod_{i=1}^n\vert t_i\vert ^{2k_jd_+}\vert 1-t_i\vert^{2k_2d_+}
\prod_{i<j}^n\vert t_i-t_j\vert^{2d_+^2}\prod_{i=1}^n
\prod_{j=1}^m\vert t_i-r_j\vert ^{-2}\cr
&\times\prod_{i=1}^m\vert r_i\vert ^{2k_jd_-}\vert 1-r_i\vert^{2k_2d_-}
\prod_{i<j}^m\vert r_i-r_j\vert^{2d_-^2}\prod_{i=1}^n
\prod_{j=4}^N\vert t_i-\tilde  z_j\vert ^{2d_+k_j}
\prod_{i=1}^m\prod_{j=4}^N\vert r_i-\tilde  z_j\vert ^{2d_-k_j}\cr
&\times \!\left\langle \!(\beta^2_1\overline \psi \psi (0)\!-\!k_1^2
\overline \xi \xi(0))
\prod_{j=4}^N\!(\beta^2_j\overline \psi\psi(\tilde  z_j)\!-\!k_j^2\overline
\xi \xi (\tilde  z_j))\prod_{i=1}^n\!\overline \xi\xi(t_i)\!\prod_{i=1}^m\!
\overline\xi\xi(r_i)\! \prod_{i=1}^s\overline \psi\psi(w_i)\!\right\rangle_0
& (4.2)\cr}
$$
Now we have several terms which give non-trivial amplitudes,
in the following we assume $m+n$ and $N+s$ even, so we have, for instance:
$$
\eqalignno{
{\cal A}^{nm}_N&=\Gamma(-s)\left( -{\pi\over 2}\right)^3\left(
{i\mu\alpha_+^2\over \pi} \right)^s(-d_+^2)^n(-d_-^2)^m \left(
\prod_{j=4}^N\beta^2_j\right) (\alpha_+)^{-2}\cr
&\times \alpha^2\prod_{j=4}^N\int d^2\tilde  z_j\prod_{i=1}^n\int d^2t_i
\prod_{i=1}^m\int d^2r_i
\prod _{i=1}^sd^2w_i\vert w_i\vert ^{2\alpha}\vert
1-w_i\vert ^{2\beta}\cr
&\times \prod_{i<j}\vert w_i-w_j\vert^{4\rho}
\prod_{i=1}^s\prod_{j=4}^N\vert w_i-\tilde  z_j\vert ^{2p_j}
\prod_{j=4}^N\vert \tilde  z_j\vert ^{2\alpha'_j}\vert
1-\tilde  z_j\vert ^{2\beta'_j)}\prod _{j<l=4}^N\vert
\tilde  z_j-\tilde  z_l\vert ^{2\rho'_{jl}}\cr
&\times\prod_{i=1}^n\vert t_i\vert ^{2\tilde  \alpha}
\vert 1-t_i\vert^{2\tilde  \beta}
\prod_{i<j}^n\vert t_i-t_j\vert^{2\tilde  \rho}
\prod_{i=1}^n\prod_{j=1}^m\vert t_i-r_j\vert ^{-2}\cr
&\times\prod_{i=1}^m\vert r_i\vert ^{2\tilde  \alpha'}
\vert 1-r_i\vert^{2\tilde\beta'}\prod_{i<j}^m\vert r_i-r_j\vert^{2\tilde\rho'}
\times \prod_{i=1}^n\prod_{j=4}^N\vert t_i-\tilde  z_j\vert ^{2\tilde
\alpha_j}\prod_{i=1}^m\prod_{j=4}^N\vert r_i-\tilde  z_j\vert ^{2\tilde
\alpha'_j }\cr
&\times \left\langle \prod_{i=1}^n\overline \xi \xi (t_i)\prod_{i=1}^m
\overline \xi \xi(r_i)\right\rangle _0\left\langle\overline \psi\psi (0)
\prod_{j=1}^s \overline \psi\psi(w_i)\prod_{j=4}^N\overline
\psi \psi (\tilde  z_j)\right\rangle _0 &(4.3)\cr}
$$
with $\alpha,\beta,\rho,\tilde \alpha,\tilde \beta,\tilde \rho,
\tilde \alpha',\tilde \beta',\tilde \rho'$ defined as before and
$$
\eqalign{
\alpha'_j&=k_1k_j-\beta_1\beta_j\quad ,
\quad \tilde  \alpha_j=d_+k_j\quad ,\quad \cr
\beta'_j&=k_2k_j-\beta_2\beta_j\quad ,
\quad \tilde  \alpha'_j=d_-k_j\quad ,\quad \cr
\rho'_{jl}&={1\over 2}(k_jk_l-\beta_j\beta_l)\quad ,\quad
p_j=-\alpha_+\beta_j\quad ,\quad 4\le j,l\le N\cr}\eqno(4.4)
$$

Using the kinematics: $k_1,k_2,\cdots,k_{N-1}\ge\alpha_0\, ,
\, k_N<\alpha_0$ it
is possible to eliminate all parameters
in terms of $\alpha,\beta,\rho$ and
$p_j (4\le j\le N-1)$:
$$\eqalign{
p_N&=-{(m+1)\over 2}-\rho(N+s+n-3)\cr
\alpha'_j&=\alpha+p_j-2\rho\quad ,\quad \beta'_j =
\beta+p_j-2\rho\quad ,\quad
4\le j\le N-1\cr
\alpha'_N&={(m-1)\over 2}+(\rho-\alpha)(N+s+n-3)
-{m\rho^{-1}\beta\over 2}\cr
\rho'_{jn}&={(m-1)\over 4}-{m\rho^{-1}\over 4}p_j+{(\rho-p_j)\over
2}(N+s+n-3)\cr
\rho'_{jl}&={1\over 2}(p_l+p_j)-\rho\cr
\tilde  \alpha&=\alpha-2\rho\quad ,\quad
\tilde  \beta=\beta-2\rho\quad,\cr
\tilde  \alpha'&=-1+{\rho^{-1}\over 2}\alpha \quad ,\quad \tilde
\beta'=-1{\beta\over 2}\rho^{-1}\cr }\eqno(4.5)
$$
where $4\le j,l\le N-1$. Using the symmetries:
$$
{\cal A}_N^{nm}(\alpha,\beta,\rho,p_1,p_2,\cdots, p_{N-1})={\cal
A}_N^{nm}(\beta,\alpha,\rho,p_1,p_2,\cdots, p_{N-1})\eqno(4.6)
$$
$$
{\cal A}_N^{nm}(\alpha,\beta,\rho,p_1,\cdots, p_{N-1})\! =\!
{\cal A}_N^{nm}(\! - \alpha  -  \beta \! + \!
{(m \! - \! 1)\over 2}-\!
P\! + \rho(N \! + n  -  s - 1),\beta,
\rho,p_1,\cdots  ,p_{N-1}\!)
\eqno(4.7)
$$
(with  $P=\sum \limits_{j=4}^{N-1}p_j$), and the large-$\alpha$ behaviour:
$$
{\cal A}_N^{nm}(\alpha\to\infty)\sim
\alpha^{1-m+2\beta+2\rho(s-N-n+3)+2P}\eqno(4.8)
$$
we have  the ansatz:
$$
{\cal A}_N^{nm}\! =\! f_N^{nm}\!(\rho,p_1,\cdots ,p_{N-1}\!)
\Delta({1\over
2}\!+\!\rho\!-\!\alpha)\Delta(\!{1\over 2}\!+\!\rho\!-\!\beta\!)
\Delta (\!1-
{m\over 2}\!+\!P\!+\!\alpha\!+\!\beta\!
+\!\rho(\!2\!+\!s\!-\!n\!-\!N\!))
\eqno(4.9)
$$
By sending $k_i\, (3\le i\le N-1)$ to zero,
which implies $p_j\to 2\rho \,
 (4\le j\le N-1)$, we can determine $f_N^{nm}(\rho,p_1,
\cdots p_{N-1})$ using:
$$
{\cal A}_N^{nm}(\alpha,\beta,\rho,k_i\to 0)=\left( -{i\pi\over
2}\right)^{N-3}{\partial ^{N-3}\over \partial _\mu}{\cal A}_3^{nm}
(k_1,k_2,k_N )\eqno(4.10)
$$
and the result for ${\cal A}_3^{nm}$ (see (3.26)). We get
$$\eqalignno{
f_N(\rho,p_1,\cdots , p_{N-1})&=
\left[ -{i\pi\over 2}\Delta({1\over
2}+\rho)\right] ^n
\left[ -{i\pi\over 2}\Delta({1\over 2}+{\rho^{-1}\over 4})
\right] ^m
\left( -{i\pi\over 2}\right)^N\cr
&\times \left[ \Delta ({1\over 2} \! - \! \rho)\right]^s
\left(
\prod_{i=4}^{N-1}\Delta ({1\over 2} \! + \! \rho \! - \! p_j)\right) \!
 \Delta \!  \left( -{(s \! + \! n \! + \! N \! - \! 4)\over 2}
- \! {m\over 4}
\rho^{-1}\right)\cr
&\times {\partial \over \partial \mu }^{N-3}
\left[ {\mu \over
2}\right]^{s+N-3}\quad . &(4.11)\cr }
$$

So, the final result for the general $N$-point function with arbitrary
screening charges can be written in a simple form:
$$\eqalignno{
{\cal A}_N^{nm}&=(s+N-3)(s+N-4)\cdots (s+1)s\left[ \mu \Delta ({1\over
2}-\rho)\right]^s\cr
&\times \left[ -{i\pi\over 2}\Delta({1\over
2}+\rho)\right] ^n \left[ -{i\pi\over 2}\Delta({1\over 2}+\rho)
\right] ^m \prod_{i=1}^N \left( -{i\pi\over 2}\right)^N\Delta({1\over
2}(1+\beta_i^2-k_i^2))&(4.12)\cr }
$$
Redefining $\Psi_{NS}\, ,\, \mu $ and
the screening charges we have our final
result:
$$
{\cal A}_N^{nm}={\partial\over \partial\mu}^{N-3}\mu^{s+N-3}\eqno(4.13)
$$

\vskip 1cm
\penalty-200
\centerline {\bf 4- Conclusion }
\vskip 1cm
 \nobreak
We have computed exactly N-point correlators in the NS sector of super
Liouville theory conformally coupled to $c\le 3/2$ supermatter in a
supercoulomb gas representation including an arbitrary number of screening
charges in the matter sector. We also generalized previous results for bosonic
amplitudes to the case including arbitrary s.c.. We have learned that in all
those cases the final N-point amplitude with $n,m$ s.c. has the same (rather
simple) form given by (4.13) above in terms of the renormalized cosmological
constant $\mu$ and the parameter $s$
which is a function of the matter central charge and the external momenta.
This confirms suspicions\ref{18,19} in that direction using a proposal of the
matrix model approach. The close connection to the bosonic amplitude
was suggested from super KP systems as well\ref{19}. This demonstration is
however more direct (see also [15] and [16]).
The similarity is striking, and the hope is that a full treatment of
the theory by the super matrix
model approach could be checked against our results.
Furthermore, the 4-point NS amplitude at $s=0$ recently obtained by Di
Francesco and Kutasov\ref{16} can be also compared with our result (in that
limit) and the check is positive. It should be stressed, however, that our
results for arbitrary $s$ permits
 to visualize the role of the barrier at $c=1$ (in the
bosonic case), since the renormalization $\mu \to \mu/\Delta(-\rho)$ is
infinity at $c=1$ (where $\rho =-1$). In the supersymmetric case, the barrier
occurs at $c=3/2$, the renormalization $\mu \to \mu/\Delta(1/2 -\rho)$
is infinity at $c=3/2$ (where $\rho =-1/2$). It would be interesting to see
whether this barrier indeed disappears for N=2 supersymmetry, and we hope to
obtain also the N=2 super correlators. The next interesting question concerns
the Ramond sector. Work in this direction is in progress. There are, however,
new difficulties in that case.
\vfill \eject
\centerline {\bf Appendix A}
\vskip .2cm
In this appendix we calculate ({\it for $n$ and $m$ even}) the following
integral:
$$
\eqalignno{
I^{nm}_M(  \alpha, \beta;  \rho)& =
\prod _{i=1}^n\int d^2t_i\vert t_i\vert ^{2  \alpha}\vert 1-
t_i\vert^{2  \beta}\prod _{i<j}^n\vert t_i-t_j\vert ^{2
\rho}\cr
&\times \prod _{i=1}^m\int d^2r_i\vert r_i\vert^{2 \alpha'}
\vert 1-
r_i\vert^{2 \beta'}\prod_ {i<j}^m\vert r_i-r_j\vert ^{2
\rho'}\prod_{i=1}^{n}\prod_{j=1}^m\vert t_i-r_j\vert^{-2}\cr
&\times \left\langle \prod _{i=1}^n\overline \xi \xi (t_i)
\prod_{i=1}^m\overline \xi \xi (r_i)
\right\rangle_0\quad , &(A.1)\cr}
$$
where $\langle \overline \xi\xi(t)\overline \xi\xi(r)\rangle_0=\vert t-r\vert
^{-2}$ and $\rho'=1/\rho\, ,\, \alpha'=-\rho'\alpha\, ,\, \beta'=-\rho'\beta$.
The above integral is the supersymmetric generalization of (B.10) of the second
reference of [12]. In order to obtain $I_{nm}$ we first notice that by making
translations $(t_i\to 1-t_i\, ,\, r_i\to 1 - r_i)$ and inversions $(r_i\to
1/r_i\, ,\, t_i\to 1/t_i)$ we have the symmetries, respectively:
$$
\eqalignno{
I_{mn}(\alpha,\beta,\rho)&=I_{mn}(\beta,\alpha;\rho) &(A.2)\cr
I_{mn}(\alpha,\beta;\rho)&=I_{mn}(m-1-\alpha-\beta-\rho(n-1),\beta;\rho)
\quad . &(A.3)\cr}
$$
Changing variables we can analyse the asymptotic behaviour of $I_{mn}$ for
large $\alpha$:
$$
I_{mn}(\alpha\to\infty,
\beta;\rho)=\alpha^{(2nm-n-m-2n\beta-2m\beta'-\rho(n-1)n-\rho'(m-1)m)}\quad
.\eqno(A.4)
$$

Another information can be used, namely that for $\rho=\rho'=-1\,
(\alpha'=\alpha\, ,\, \beta'=\beta)$ the integral must be a function of $n$ and
$m$ through the combination $n+m$. It is not difficult to check (using
Stirling's formula $\Gamma (\alpha+c)\sim \alpha^c\Gamma(\alpha)$, for
large-$\alpha$), that the following Ansatz is consistent with all requirements
   given
above:
$$\eqalign{
&I_{mn}(\alpha,\beta;\rho)=C_{mn}(\rho)\cr
&\times \prod _{i=0}^{{n\over 2}-1}\Delta (1+ \alpha + i
\rho )\Delta(1+  \beta +i \rho)\Delta(m-\alpha
- \beta + \rho (i-n+1))\cr
&\times \prod _{i=1}^{{n\over 2}}\Delta ({1\over 2}\! +\! \alpha
\!+\! (i \! - \! {1\over 2} )
\rho )\Delta({1\over 2}\! +\!\beta \! + \! (i\! -\! {1\over 2}) \rho)
\Delta(-{1\over 2}\! -\!  \alpha \! +\! m \! -  \beta \! +
\!  \rho (i\! -\! n\! +\! {1\over 2}))\cr
&\times \prod _{i=0}^{{m\over 2}-1}\Delta (1+ \alpha' - {n\over 2}
+ i \rho ')\Delta(1 - {n\over 2} +   \beta' +i
\rho' )\Delta( {n\over 2} - \alpha' -\beta '+
\rho' (i-m+1))\cr
&\times \prod _{i=1}^{{m\over 2}}\Delta ({1\over 2}\!
-\!  {n\over 2}\!  +\!  \alpha '\! +\! (i\! -\! {1\over 2} )
\rho' )\Delta({1\over 2}\! -\! {n\over 2}\! +\! \beta' \! +\!
(i\! - \! {1\over 2}) \rho')
\Delta(\! -{1\over 2}\! + {n\over 2}\! -\! \alpha'\! -\!
\beta'\! +\! \rho'(i\! -\! m \! + \! {1\over 2}))\cr}\eqno(A.5)
$$

The coefficient $C_{mn}(\rho)$ can be basically obtained by noticing that the
integral $I_{mn}$ reduces to a known integral when $m$ or $n$ vanishes:
$$
\eqalignno{
I_{0n}&=(-)^{n/2}(n!!)J^{n/2}(\alpha,\beta,\gamma=-1/2,\rho/2) &(A.6)\cr
I_{m0}&=(-)^{m/2}(m!!)J^{m/2}(\alpha',\beta',\gamma=-1/2,\rho'/2) &(A.7)\cr}
$$
We have calculated the integral $J^m(\alpha,\beta;\rho)$ in Ref. [15]. For the
reader's convenience we give the result:
$$
\eqalign{
&J^m(\alpha,\beta;\gamma;\rho)=\cr
&{\pi^{2m}\over 2^m}m! \left[ \Delta\left( -(\gamma
+\rho) \right) \right]^{2m}\prod_{i=1}^m\Delta \left(
1+2(\gamma+i\rho)\right) \Delta \left( 1+\gamma+(2i-1)\rho\right)\cr
&\times\prod_{i=0}^{m-1}\Delta( 1+\alpha+2i\rho) \Delta( 1+\beta
+2i\rho) \Delta(-1-\alpha-\beta-2\gamma+(2i-4m+2)\rho) \cr
&\times\!\prod_{i=1}^m\!\Delta(\! 1\!+\!\alpha\!+\!\gamma\!+\!
(2i\!-\!1)\rho) \Delta( 1\!+\!\beta\!+\!\gamma\!+
(2i\!-\!1)\rho) \Delta( -1\!-\!\alpha\!-\!\beta\!-
\!\gamma\!+(2i\!-\!4m\!+\!2)\rho)\cr }\eqno(A.8)
$$
Using the above result in eq.(A.6) and (A.7) with the Ansatz (A.5) we obtain
$C_{mn}(\rho)$:
$$
\eqalign{
&C_{nm}(\rho)=
(-)^{{n+m\over 2}}{\pi^{n+m}\over 2^{n+m}}n!m!
\left( -{ \rho\over2}\right) ^{-2nm}
\left[\Delta\left( {1\over 2}-{ \rho\over 2}\right)\right] ^n
\left[\Delta\left( {1\over 2}-{\rho'\over 2}\right)\right] ^m
\cr
&\times \prod _1^{n\over 2}\Delta (i \rho)\Delta\left( {1\over
2}+\rho\left( i-{1\over 2}\right) \right) \prod _1^{m\over 2}
\Delta (i \rho' -{n\over 2})\Delta\left( {1\over 2}-{n\over 2}
- \rho'\left( i-{1\over 2}\right) \right) \cr}\eqno(A.9)
$$
which determines $I_{mn}(\alpha,\beta;\rho)$ completely.

\vskip 1cm
 {\bf Acknowledgments}
\vskip .3cm
The work of K.H. (contract \# 90/1799-9) and D.D. (contract \# 90/2246-3) was
supported by FAPESP while the work of E.A. and M.C.B.A. is partially supported
by CNPq.

\vskip 1cm
\penalty-100
\centerline {\bf References}
\def\refer[#1/#2]{ \item{#1} {{#2}} }
\vskip .5cm
\nobreak
\refer[[1]/E. Br\'ezin and V. A. Kazakov, Phys. Lett.  {\bf B236}
(1990)144; M. R. Douglas and S. H. Shenker, Nucl. Phys.  {\bf B335}(1990)635;
D. J. Gross and A. A. Migdal, Phys. Rev. Lett. {\bf 64}(1990)127.]

\refer[[2]/A. M. Polyakov, Phys. Lett.  {\bf B103}(1981)207.]

\refer[/A. M. Polyakov, Phys. Lett.  {\bf B103 }(1981)211.]

\refer[[3]/T. L. Curtright and C. B. Thorn, Phys. Rev. Lett. {\bf 48},
(1982)1309; E. Braaten, T. L. Curtright and C. B. Thorn, Phys. Lett.  {\bf
B118}(1982)115; Ann. Phys. {\bf 147}(1983)365; E. Braaten, T. L.
Curtright, G. Gandour and C. B. Thorn, Phys. Rev.
Lett. {\bf 51}(1983)19; Ann. Phys. {\bf 153}(1984)147; J. L. Gervais and A.
Neveu, Nucl. Phys.  {\bf B199}(1982)50;  {\bf B209}(1982)125;  {\bf B224}
(1983)329;  {\bf B238}(1984)123,396; E. D'Hoker and R. Jackiw, Phys.
Rev.  {\bf D26}(1982)3517; T. Yoneya, Phys. Lett.  {\bf B148}(1984)111.]

\refer[/N. Seiberg, Lecture at 1990 Yukawa Int. Sem. Common Trends in Math.
and Quantum Field Theory, and Cargese meeting Random Surfaces, Quantum Gravity
and Strings, May 27, June 2, 1990; J.
Polchinski, Strings '90 Conference, College Station, TX, Mar 12-17, 1990, Nucl.
 Phys. {\bf B357}(1991)241.]

\refer[[4]/M. Goulian and M. Li, Phys. Rev. Lett. {\bf 66}(1991)2051.]

\refer[[5]/N. Sakai and Y. Tanii, Tokyo Institute of Technology preprint
TIT/HEP-169 (STUPP-91-116) (1991).]

\refer[[6]/P. Di Francesco and D. Kutasov, Phys. Lett.  {\bf B261}(1991)385.]

\refer[[7]/Vl. S. Dotsenko,  Paris VI preprint PAR-LPTHE 91-18 (1991).]

\refer[[8]/Y. Kitazawa,  Harvard Univ. preprint HUTP-91/A013 (1991).]

\refer[[9]/A. Gupta, S. P. Trivedi and M. B. Wise, Nucl. Phys.  {\bf B340}
(1990)475.]

\refer[[10]/M. Bershadsky and I. R. Klebanov, Phys. Rev. Lett. {\bf 65},
3088 (1990); Harvard Univ. preprint HUTP-91/A002 (PUPT-1236) (1991)
; A. M. Polyakov, Mod. Phys. Lett.  {\bf A6}(1991)635.]

\refer[[11]/J. F. Arvis,
Nucl.\ Phys.\  {\bf B212}(1983)151;  {\bf B218}(1983)309; O. Babelon,
Nucl. Phys. {\bf B258}(1985)680; Phys. Lett. {\bf 141B}(1984)353; T.
Curtright and G. Ghandour, Phys. Lett. {\bf B136}(1984)50.]

\refer[[12]/Vl. S. Dotsenko and V. Fateev, Nucl. Phys.  {\bf B240}
(1984)312;  {\bf B251}(1985)691.]

\refer[[13]/J. Distler, Z. Hlousek and H. Kawai, Int. J. Mod. Phys. {\bf A5}
(1990)391.]

\refer[[14]/E. Martinec, Phys. Rev. {\bf D28}(1983)2604.]

\refer[[15]/E. Abdalla, M.C.B. Abdalla, D. Dalmazi and K. Harada, ``
Correlation functions in super Liouville theory", preprint IFT-P.029/91
(1991).]

\refer[[16]/P. Di Francesco and D. Kutasov, Princeton Univ. preprint PUPT-1276
(1991).]

\refer[/K. Aoki and E. D'Hoker, preprint UCLA/91/TEP/33(1991).]

\refer[[17]/F. David, Mod. Phys. Lett. {\bf A3}(1988)1651; J. Distler and
H. Kawai, Nucl. Phys.  {\bf B321}(1989)509.]

\refer[[18]/L. Alvarez-Gaum\'e and J. L. Ma\~nez, Mod. Phys. Lett.
{\bf A6}, 2039 (1991).]

\refer[[19]/P. Di Francesco, J. Distler and D. Kutasov, Mod. Phys. Lett.
{\bf A5}, 2135 (1990).]


\end